\documentclass[twocolumn,showpacs,preprintnumbers,amsmath,amssymb]{revtex4}
\usepackage{graphicx}
\usepackage{dcolumn}
\usepackage{bm}

\begin{document}

\title{Transverse momentum spectra of hadrons in pp, pA and AA
collisions}

\author{Bedangadas Mohanty}
\medskip

\affiliation{Variable Energy Cyclotron Centre, Kolkata 
   700064, India}

\date{\today}

\begin{abstract}
The transverse momentum spectra of the produced hadrons 
have been compared to a model which is based on the assumption 
that a nucleus-nucleus collision is a superposition of 
isotropically decaying thermal sources at a given freeze out 
temperature. The freeze-out temperature in nucleus-nucleus collisions is
fixed from the inverse slope of the transverse momentum spectra
of hadrons in nucleon-nucleon collision. The successive collisions 
in the nuclear reaction leads to gain in transverse momentum, 
as the nucleons propagate in the nucleus following a random walk pattern. 
The average 
transverse rapidity shift per collision is determined from the 
nucleon-nucleus collision data. Using these information
we obtain parameter free result for the transverse momentum 
distribution of produced hadrons in nucleus-nucleus collisions. 
It is observed that such a model is able to explain the transverse 
mass spectra of produced pions at SPS energies. However it fails 
to satisfactorily explain the transverse mass spectra of kaons
and protons. This indicates the presence of collective effect which
cannot be accounted for by the initial state collision broadening 
of transverse momentum of produced hadrons, the basis of random walk model.

\end{abstract}

\maketitle

\section{INTRODUCTION}

\begin{figure}
\begin{center}
\vspace{-1cm}
\includegraphics[scale=0.4]{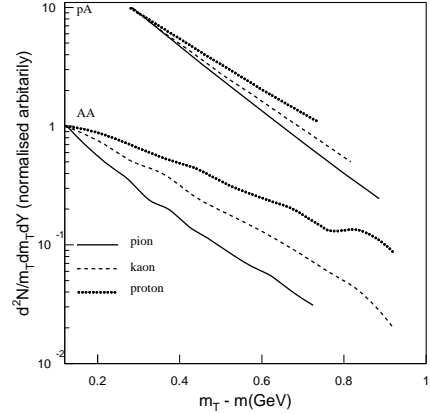}
\caption{Experimental transverse mass spectra of produced pions, 
kaons and protons for nucleon-nucleus (pA) and nucleus-nucleus (AA) 
collisions showing the effect $p_T$ broadening with mass of hadrons. 
The spectra's have been arbitrarily normalized to 10 and 1
at the lowest momentum bin for pA and AA collisions respectively.
The pA data corresponds to AGS energy while the AA data to SPS
energy.
}
\label{flow_spectra}
\end{center}
\end{figure}

In high energy heavy-ion collisions it is observed that the average
transverse momenta of the produced hadrons depends strongly on the
mass of hadrons~\cite{tflow,bdm}. Further it is found to be substantially 
larger than in nucleon-nucleon collisions at a given energy. 
One of the possible physical effect responsible for this is 
transverse flow~\cite{heinz}. 
The heavier the mass of the particle more is the 
gain in momentum and hence more transverse flow. This has been shown
by extracting the  effective temperature (effective because it includes
the true freeze-out temperature and the transverse flow) 
from transverse momentum ($p_T$) distribution of various hadrons. 
It is observed that protons 
have a larger effective temperature than kaons, which in turn has a
larger effective temperature than pions. The success of hydrodynamic
models in explaining the transverse momentum distributions of various
hadrons at least upto lower momentum, is an indication of presence 
of radial flow~\cite{radial}. 
Fig.~\ref{flow_spectra} depicts this effect for AA collisions at SPS energies. 
However similar effect is also observed in nucleon-nucleus (pA) 
collisions (also shown in Fig.~\ref{flow_spectra}). 
Hence broadening of $p_T$ even in pA collisions, 
where one does not expect any collective effect like radial flow, indicates
there exists a ``normal'' $p_T$ broadening observed in all heavy-ion 
reactions. This needs to be understood properly before making any 
quantitative conclusions regarding the radial flow. 
Such broadening of $p_T$ is thought to arise due to successive 
collisions in nuclear reactions. It is basically an initial state collision
broadening of $p_T$. 
This effect in high $p_T$ (typically above 2 GeV) is referred to as 
Cronin effect~\cite{cronin}.  
Apart from the commonly produced hadrons, the Drell-Yan dileptons and
quarkonium resonances also show in pA collisions a broadening of
$p_T$ compared to that observed in pp collisions. The virtual photons
in Drell-Yan production do not interact in the nuclear medium and the
quarkonia leave the medium before the collective effects develop. This
makes the case for the $p_T$ broadening as an initial state effect 
stronger~\cite{leonidov,satz}.

In ~\cite{leonidov,jane}, it has been
shown that such an effect is able to explain the transverse mass ($m_T$)
distribution of produced hadrons at SPS energies. Thereby questioning the
presence of true collective effects (such as radial flow) in relativistic
heavy ion collisions.
The aim of the present work is to have a detail study of the effect of 
$p_T$ gain through successive collisions in nuclear reaction within
the framework of a random walk model~\cite{leonidov,jane} 
(discussed in the next section).
In this paper we systematically analyze 
the available pp, pA and AA data and come to the conclusion that 
collective effects does exists in relativistic heavy-ion collisions.  

The paper is organized as follows. In the next section we discuss the 
random walk model. In section III we fix the freeze-out temperature 
for AA collisions from the inverse slope of the $p_{T}$ spectra
from the pp collisions. Section IV deals with the estimation of the
gain in average transverse momentum per collision from pA collision data.
We compare the results of the random walk model calculation with the 
SPS AA collision data in section V. In section VI we give a discussion
of the results. Finally a summary of the work is given in section VII.

\section{RANDOM WALK MODEL}

Consider the nucleus-nucleus collision to be a superposition of 
nucleon-nucleon collision. In this picture, let us assume that in 
each successive interactions of a nuclear collision one creates a fire ball 
just like that formed in a nucleon-nucleon collision. 
If a nucleon starts with zero $p_T$, after the first collision the next 
one will generally occur 
at some non-vanishing transverse velocity. Thus there is a gain 
in transverse momentum through successive collisions. The propagation of 
nucleon through successive collisions is assumed to follow a 
$random~walk$ pattern. It is of interest to find if such a model can explain
the $p_T$ spectra of produced hadrons in AA collisions and more specifically
if it can account for the observed $p_T$ broadening with increase in 
mass of the hadrons.
The random walk pattern is modeled by a Gaussian \cite{leonidov} in the
present calculation as,
\begin{equation}
f_{pA}(\rho ) = \left[ {4\over \pi\delta^2_{pA}}\right]^{1/2}\exp
(-\rho^2/\delta^2_{pA}),
\label{eq1}
\end{equation}
where $\rho$ is the transverse rapidity and
\begin{equation}
\delta_{pA}^2 = (N_A-1)\delta^2,
\label{eq2}
\end{equation}
denotes the kick per collision $\delta$ as determined from pA
interactions. $N_A$ is the number of nucleons which the incident
proton encounters on its path through the target nucleus. It is given
by~\cite{leonidov}
\begin{equation}
N_A \sim (3/4)(2\pi r_{0}^{2} R_{A}) n_{0},
\label{eq3}
\end{equation}
where $r_{0}~\sim~0.8$ fm is the nuclear radius, $R_A$ (= 1.12 $A^{1/3}$ fm)
is the radius of the nucleus and $n_{0}$ = 0.17 $fm^{3}$ is the nuclear 
density.

The corresponding distribution for an $AB$ collision can be shown to
of the form,
\begin{equation}
f_{AB}(\rho ) = \left[ {4\over \pi\delta^2_{AB}}\right]^{1/2}\exp
(-\rho^2/\delta^2_{AB}),
\label{eq4}
\end{equation}
with
\begin{equation}
\delta_{AB}^2 = (N_A+N_B-2)\delta^2.
\label{eq5}
\end{equation}

The final expression for the transverse mass distribution is given 
as~\cite{jane}
\begin{widetext}
\begin{equation}
\left( {dN\over dy m_Tdm_T}\right)_{y=0}
= {gV\over 2\pi^2}
  \left[ {4\over \pi\delta^2_{AB}}\right]^{1/2}\int d\rho\exp
(-\rho^2/\delta^2_{AB})
m_TI_0 \left( {p_T\sinh\rho\over T} \right)
          K_1 \left( {m_T\cosh\rho\over T} \right).
\label{eq6}
\end{equation}
\end{widetext}
Where $I_{0}$ and $K_1$ are modified Bessel functions, T is the freeze-out
temperature and 
$m_{T}~=~\sqrt{p_{T}^{2} + m^{2}}$ 
is the transverse mass.

It should be noted that the volume in the Eqn.~\ref{eq6} refers to the
volume of the system as observed in a $pp$ collision, since each
collision in the random walk produces a $pp$ type of fireball. 
If we now introduce a boost-invariant distribution of fireballs along
the longitudinal rapi\-di\-ty axis, we finally obtain by integrating
over the fireball distributions
\begin{widetext}
\begin{eqnarray}
\left( {dN\over dy m_Tdm_T}\right)_{y=0}
&=& {gV\over 2\pi^2}
  \left[ {4\over \pi\delta^2_{AB}}\right]^{1/2}\int d\rho\exp
(-\rho^2/\delta^2_{AB})\nonumber\\
& & \int_{-Y_L}^{Y_L} dY
m_T\cosh YI_0 \left( {p_T\sinh\rho\over T} \right)
          K_1 \left( {m_T\cosh Y\cosh\rho\over T} \right).
\label{eq7}
\end{eqnarray}
\end{widetext}

The above expression has two parameters, T and $\delta$. 
The temperature is assumed to be same for the whole fireball.
The temperature is fixed from the pp collisions and $\delta$ is 
calculated from pA collisions as mentioned before.
Fixing these two parameters from pp and pA collisions we will attempt to
describe the data for AA collisions.

\section{TRANSVERSE MOMENTUM SPECTRA IN pp COLLISIONS}
\begin{figure}
\begin{center}
\vspace{-1cm}
\includegraphics[scale=0.4]{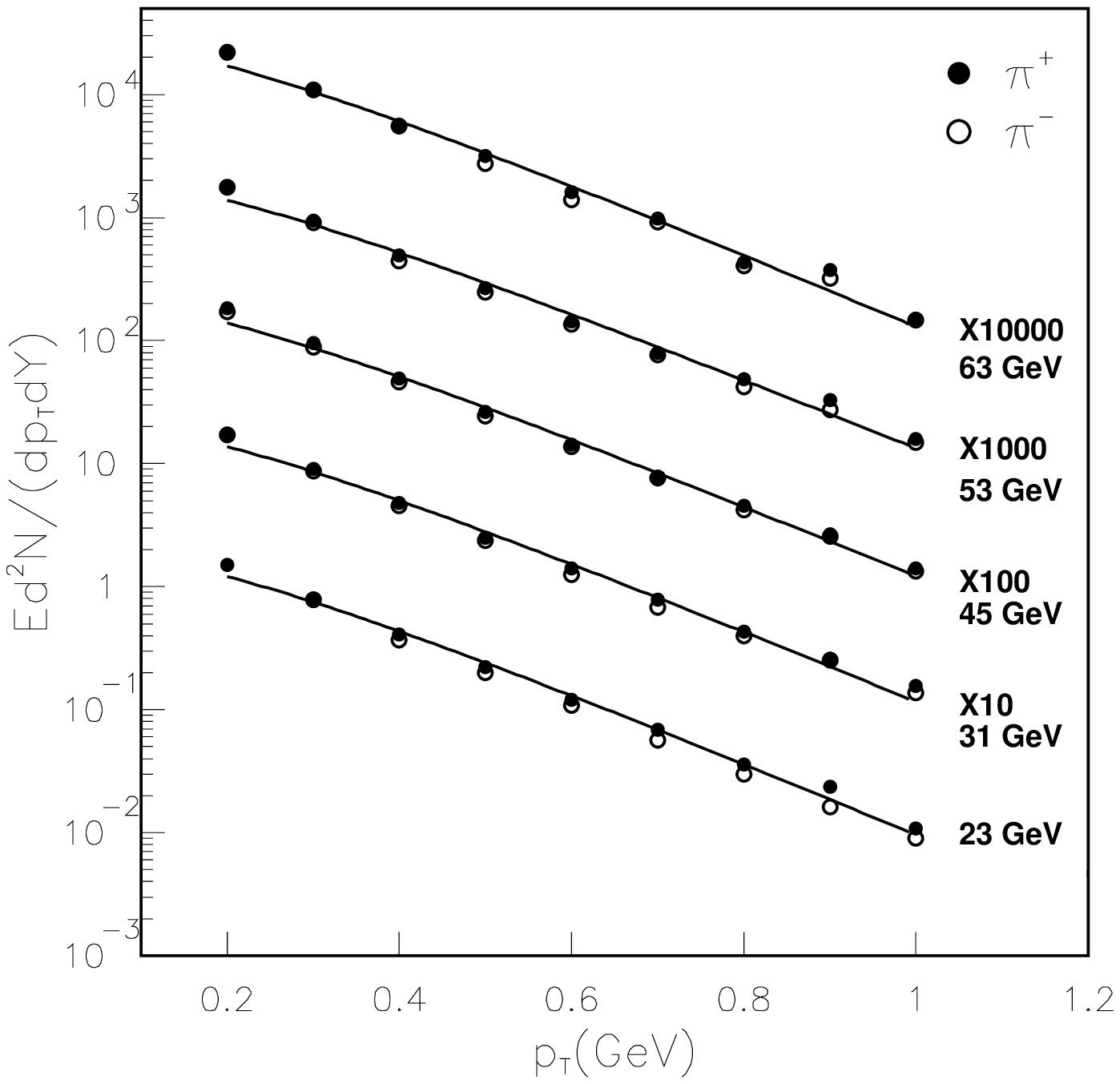}
\includegraphics[scale=0.4]{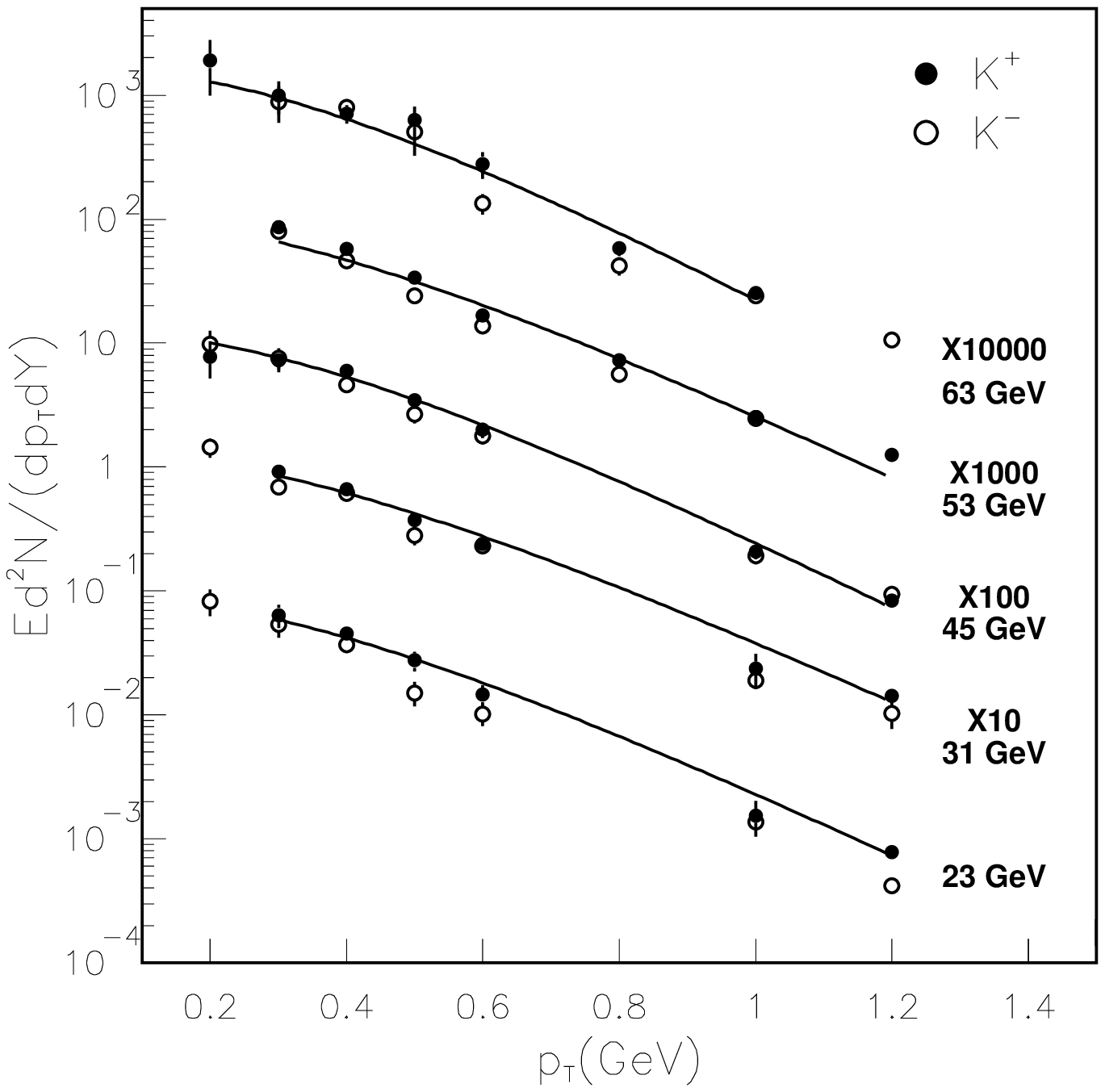}
\includegraphics[scale=0.4]{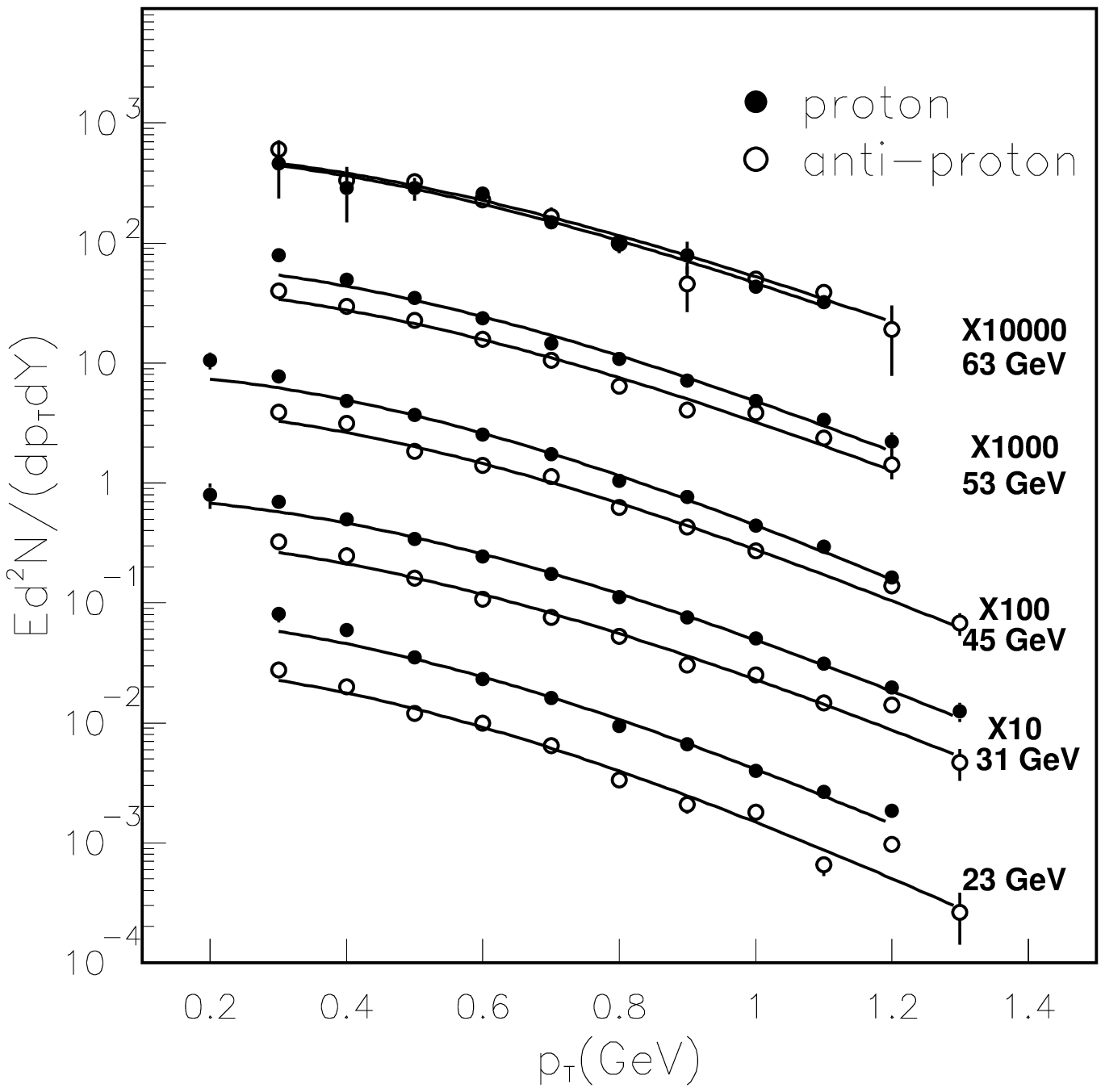}
\caption{
Transverse momentum spectra of pions, kaons, protons
and anti-protons for various CMS energies. The data has been 
scaled appropriately for clarity of presentation. The solid lines corresponds
to results obtained from Eqn.~\ref{eq8}. The fits for $\pi^{-}$
and $K^{-}$ are not shown in the figure.
}
\label{pp_spectra}
\end{center}
\end{figure}
\begin{figure}
\begin{center}
\vspace{-1cm}
\includegraphics[scale=0.4]{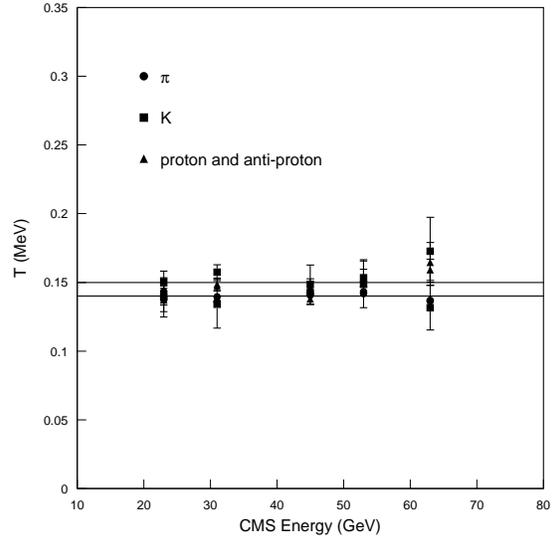}
\caption{ Temperature obtained by fitting the transverse momentum
spectra presented in Fig.~\ref{pp_spectra}, for various hadrons.
The solid lines corresponds to 140 and 150 MeV. The error bars corresponds
to the error on temperature due to the fitting.
}
\label{pp_temp}
\end{center}
\end{figure}

As discussed in the previous section, we would fix the freeze-out 
temperature  for the AA collisions from the inverse slope of the 
hadronic transverse momentum spectra originating from pp collisions.
At high
energies the transverse momentum spectra of produced hadrons can
be described by the following expression,
\begin{equation}
(\frac{d^2 N}{dydp_{T}^{2}})^{pp}~\sim~const.~m_{T} K_{1}(m_{T}/T).
\label{eq8}
\end{equation}

Using the expression given in Eq.~\ref{eq8}, we fitted the experimental
data of pp collisions of pions, kaons and protons at different centre-of-mass 
(CMS) energies
varying from 23 GeV to 63 GeV~\cite{isr}. 
The results are shown in Fig.~\ref{pp_spectra}.
The extracted temperature for  the various hadron species are shown in 
Fig.~\ref{pp_temp}. The error bars shown corresponds to the error on the
slope parameter due to the fitting procedure adopted here.
One observes that there is very little variation in T for all the produced 
hadrons in the range of CMS energy for which data is available and analyzed.
We will consider two values of temperature for our study of AA data, one 
corresponding to 140 MeV and the other corresponding to 150 MeV.

\section{TRANSVERSE MOMENTUM SPECTRA IN pA COLLISIONS}
\begin{figure}
\begin{center}
\vspace{-1cm}
\includegraphics[scale=0.4]{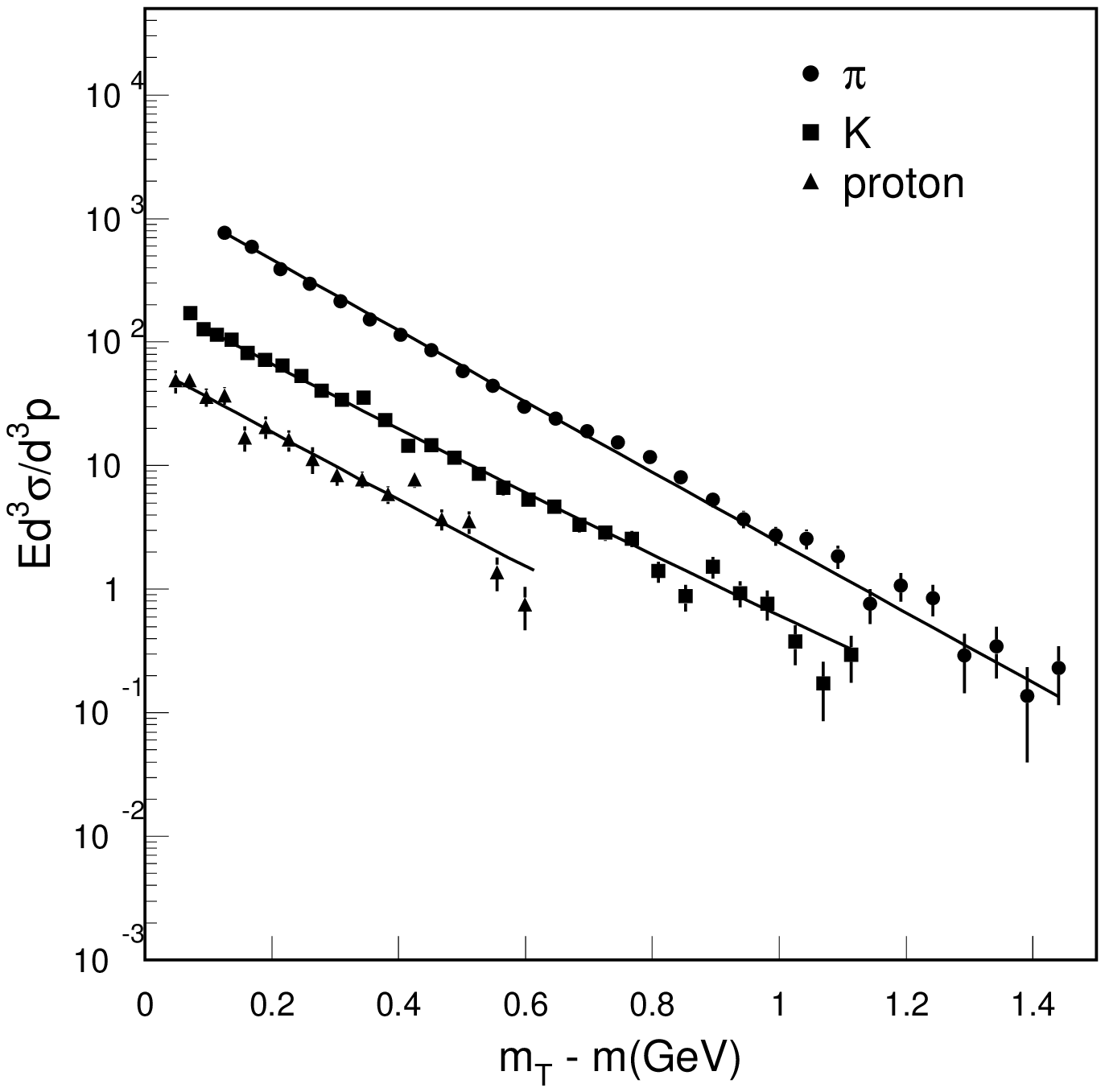}
\includegraphics[scale=0.4]{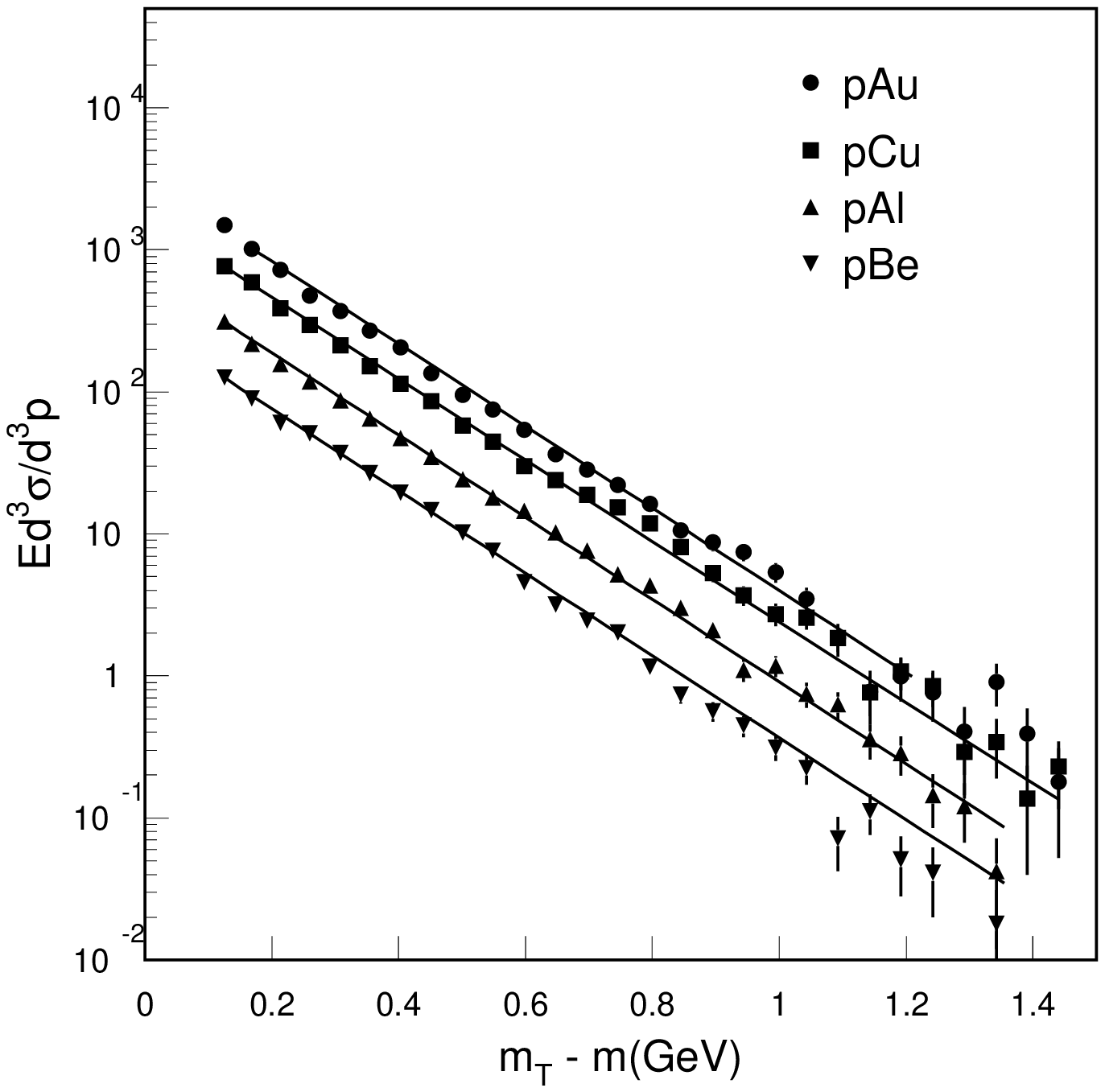}
\caption{
The top figure shows the transverse momentum spectra for 
pions, kaons, protons and anti-protons at 14.6 AGeV for
p+Be reaction. The bottom figure shows the
corresponding distribution for pions for various targets.
The solid lines corresponds to results obtained from random walk model. 
}
\label{pa_spectra}
\end{center}
\end{figure}
Having fixed the the freeze-out temperature for AA collisions from the 
inverse slope of $p_T$ spectra of produced hadrons in pp collisions, 
we now try to get the average transverse
rapidity shift per collision, $\delta$. This we will determine empirically
from transverse mass spectra of the produced hadrons in pA collisions.
To get a proper understanding of this parameter, 
we will study pA data for various available beam energies, different species
of produced hadrons and several targets. Once this
parameter is fixed, along with the temperature fixed from pp collisions,
we will be able to predict the initial state transverse momentum broadening
in AA collisions as has been discussed in the section II. The results will be 
presented in the next section.

\begin{figure}
\begin{center}
\vspace{-1cm}
\includegraphics[scale=0.5]{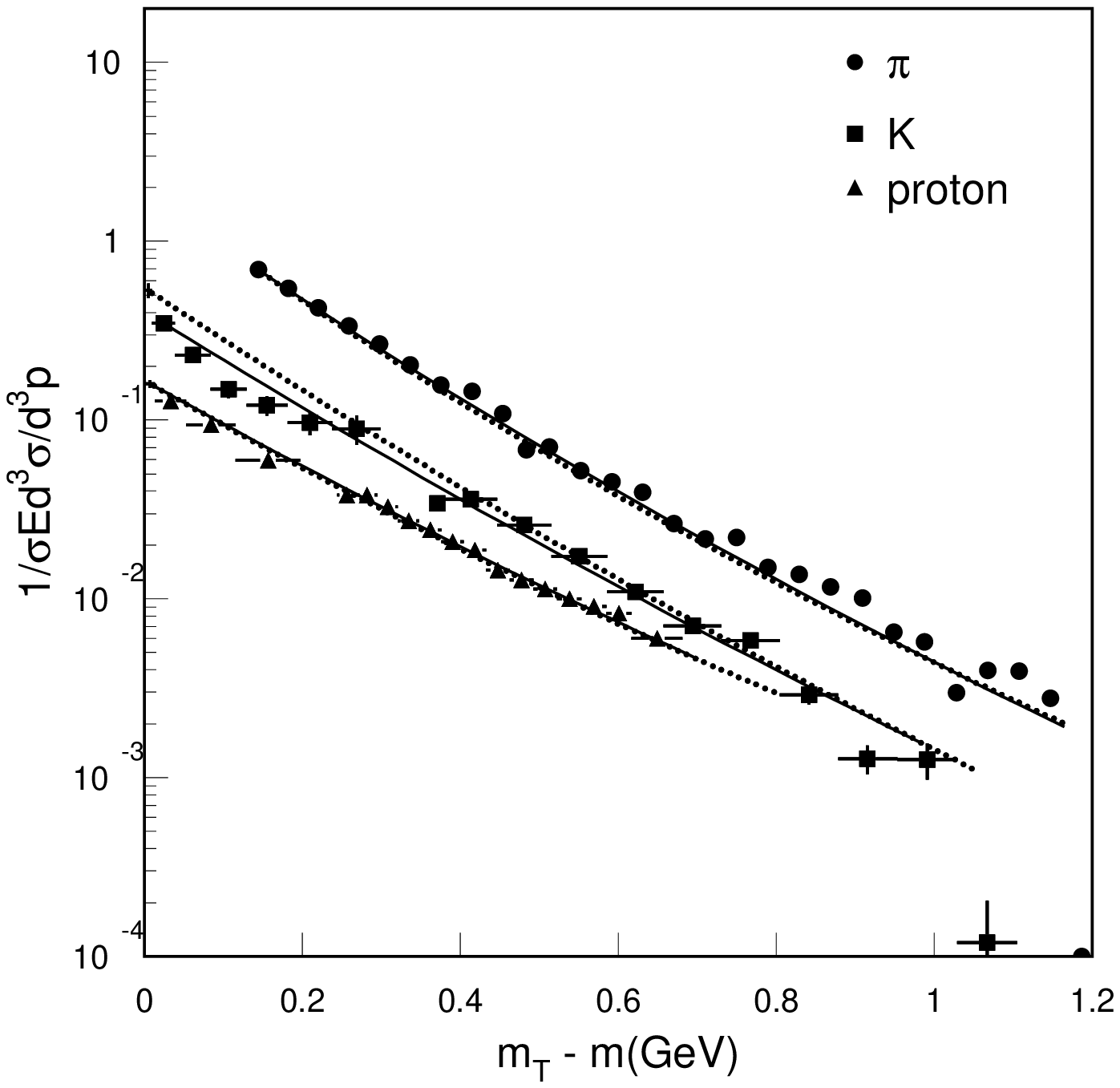}
\caption{Transverse momentum spectra for 
pions, kaons, protons and anti-protons at 450 AGeV for
p+Pb reaction. 
The solid lines corresponds to results obtained from random walk model
with T = 150 MeV and the dotted line corresponds to T = 140 MeV. 
}
\label{ppb_spectra}
\end{center}
\end{figure}
The transverse momentum spectra for different species of produced hadrons
for pA collisions can be obtained using the Eqn.~\ref{eq7}. Where 
$\delta_{pA}$ is the only parameter, with T being fixed from pp collisions.
The results are shown in Figs.~\ref{pa_spectra} and~\ref{ppb_spectra}. 
Fig.~\ref{pa_spectra} shows the comparison of transverse mass distribution
for pion, kaon and proton at 14.6 GeV for various targets~\cite{e802} with
those obtained from model calculations.
One finds that the data is reasonably well explained by the model 
calculations. The values of $\delta_{pA}$ obtained for pion, kaon and proton
are 0.05, 0.25 and 0.25 respectively. The corresponding $\delta$ values for
the p+Cu collisions are obtained using Eqn.~\ref{eq2} to be
0.04, 0.22, 0.22 for pions, kaons and protons respectively. Also shown
in Fig.~\ref{pa_spectra} is the transverse mass spectra for pions at 14.6 GeV
for various targets. We find that there is very little atomic mass dependence
of $\delta_{pA}$. It was found to be 0.05 for the various targets shown in the
figure. However one observes from Eqn.~\ref{eq5}, the average transverse
rapidity shift per collision, $\delta$ decreases with increase in atomic
mass as we go from beryllium to gold target. Fig.~\ref{ppb_spectra} shows
the transverse mass spectra for pion, kaon and proton for p+Pb 
collisions  at 450 AGeV~\cite{na44}. 
The solid lines are the result of model calculation
taking T = 150 MeV. The resulting $\delta_{pA}$, which satisfactorily fits the
data as shown, turns out to be of the order of 0.45 for three hadronic 
species.  This is a clear increase from the lower energy value, indicating
that $\delta_{pA}$ has an energy dependence. The corresponding value 
$\delta_{pA}$ for a
T = 140 MeV (dotted lines) 
lies between 0.35-0.4. 

For the description of the AA data at SPS, we will consider two values
of temperature, of 140 MeV and 150 MeV fixed from the pp data. Since the
collisions at SPS are lead on lead target, we will use $\delta_{pPb}$
values as discussed above for pPb collisions at 450 AGeV. We will also
consider one case with $\delta_{pBe}$ at 14.6 AGeV in order to get an idea
of the effect at lower energies.

\section{TRANSVERSE MOMENTUM SPECTRA IN AA COLLISIONS}
\begin{figure}
\begin{center}
\vspace{-1cm}
\includegraphics[scale=0.4]{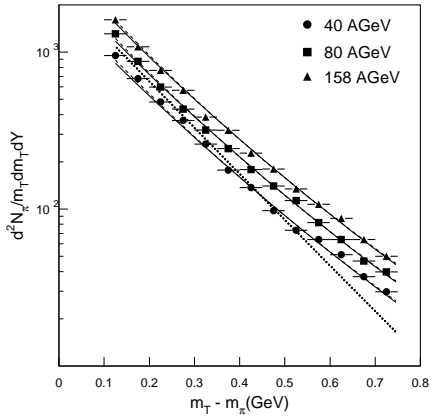}
\includegraphics[scale=0.4]{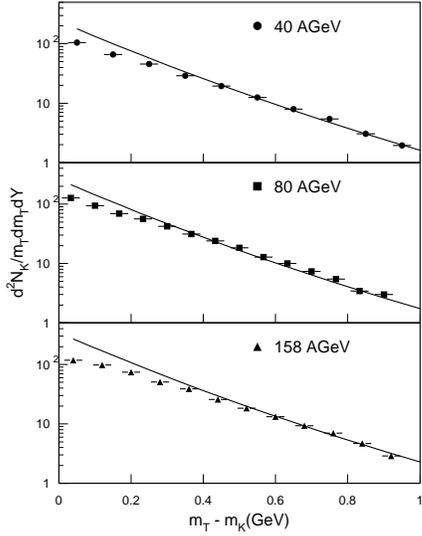}
\includegraphics[scale=0.4]{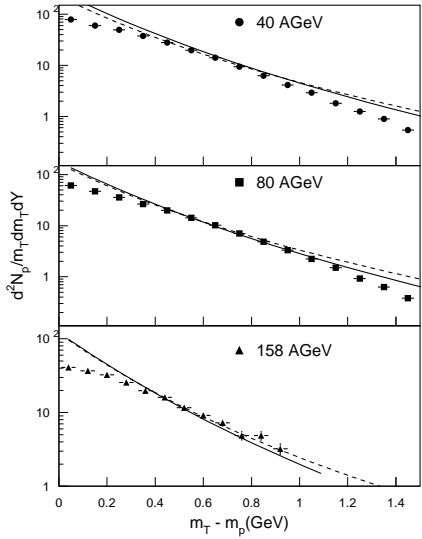}
\caption{Transverse mass spectra for pions, kaons and protons
at 40, 80 and 158 AGeV beam energy. The solid lines corresponds to calculations
from the random walk model with T = 150 MeV. The dashed lines 
corresponds to calculations with T = 140 MeV. The $\delta$ parameter
is set from pPb collision data at 450 AGeV.
The dotted line
shown for 40 AGeV pion spectra corresponds to similar calculation
at T = 150 MeV and $\delta$ obtained from pBe data at 14.6 GeV. 
}
\label{aa_spectra}
\end{center}
\end{figure}

Having fixed the temperature from the pp spectra and knowing the
value of $\delta$ from the available pA data, we now try to see
if the random walk model is able to explain the transverse mass
spectra of pions, kaons and protons at SPS energies.
The results of the model calculation using Eqn.~\ref{eq7} along with
the available data~\cite{na49} are shown in Fig.~\ref{aa_spectra}. 

The results for pions are shown in top panel of Fig.~\ref{aa_spectra}. 
The lines corresponds to the random walk model calculations for two different
temperatures of 140 MeV and 150 MeV. One observes that the model is
able to explain the pion spectra satisfactorily over the energy
range of 40 AGeV to 158 AGeV for both 140 and 150 MeV temperatures. 
The $\chi^{2}/ndf$ being less than 1.0
for all the cases. Taking a $\delta$ value corresponding to pA results
from AGS energies (14.3 AGeV) fails to explain the data. This is shown
as dotted line for 40 AGeV beam energy. Similar results
are obtained for higher beam energies and not shown in the figure for 
clarity of presentation.It may be
mentioned that the absolute 
normalisation is adjusted to get the best possible fit.
For the kaons (middle panel of Fig.~\ref{aa_spectra}), 
one finds that the model fails to explain the lower
transverse momentum region. Further one notices that the $\chi^{2}/ndf$
worsens as we go to higher beam energies. It varies from 3.5 at 40AGeV to
10 at 158 AGeV. There is not much difference between the results for the
model calculation for temperature of 140 and 150 MeV, hence the spectra
corresponding to 140 MeV is not shown in the figure. The model also 
fails to explain the observed transverse mass spectra of protons at both
lower and higher transverse momentums. 
The results for two different temperatures
of 140 MeV and 150 MeV are shown in the bottom panel of 
Fig.~\ref{aa_spectra}. The 
$\chi^{2}/ndf$ for the three cases lies between 3.5 to 7.5. 

\section{DISCUSSION}
The results indicate the following : (i) the random walk model fails to
explain the transverse mass spectra as the mass of the hadron increases 
where the effect of possible transverse flow will be more and  
(ii) the $\chi^{2}/ndf$ indicates that the disagreement  of the model 
calculation with data for kaons and protons increases with increase in 
beam energy.  In this section we discuss the limitations of the model
and try to estimate the relative contribution of initial state $p_T$
broadening and transverse flow.

\subsection{Random walk pattern}
\begin{figure}
\begin{center}
\vspace{-1cm}
\includegraphics[scale=0.4]{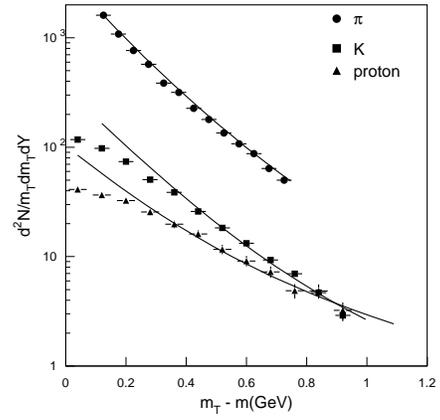}
\caption{
Transverse mass spectra for kaons and protons at 158 AGeV Pb+Pb
Collisions. The solid line corresponds to results obtained from random
walk model with the random pattern following a Lorentzian distribution. 
The model parameters T and $\delta$ are not fixed from pp 
and p+Pb collisions respectively. The normalization is 
adjusted to give the best possible agreement with the experimental data.
}
\label{linear_spectra}
\end{center}
\end{figure}
The random walk pattern was chosen to be a Gaussian (Eqn.~\ref{eq1}). 
In principle this can be any other type of statistical distribution, 
such as a Lorentzian. We have studied the sensitivity of the result
to such a distribution. We find that the p+Pb data for pion, kaon and 
proton is well explained by a $\delta_{pPb}$ of 0.2, 0.15 and 0.35
respectively. The values are lower compared to those obtained considering
a Gaussian random walk pattern. Extrapolating these values to Pb+Pb 
collisions and taking the value of $T$ to be 150 MeV, we obtain the
$m_T$ spectra for the available SPS data at the highest
energy. The results
are shown in Fig.~\ref{linear_spectra}. We observe that the model fails
to explain the data satisfactorily, specially at low $m_T$ for kaons
and protons. However it agrees fairly well with the observed pion spectra.
It seems that the results are insensitive to the choice of random walk pattern.

\subsection{Parameters of random walk model}
One of the disadvantage of the present work is the absence of the pp 
and pA data in literature corresponding to the exact SPS energies for
which AA collision data is available. However
one must mention that the near constant temperature exhibited by the 
available pp data for a wide range of CMS energy more or less fixes the 
temperature parameter of the model. 
Contrary to this one observes a considerable variation of
$\delta$ with beam energy. In view of these, it is
obvious to ask, if the SPS kaon and proton transverse mass spectra 
can be explained within the framework of random walk model for arbitrary 
values of temperature and $\delta$. 
\begin{figure}
\begin{center}
\vspace{-1cm}
\includegraphics[scale=0.4]{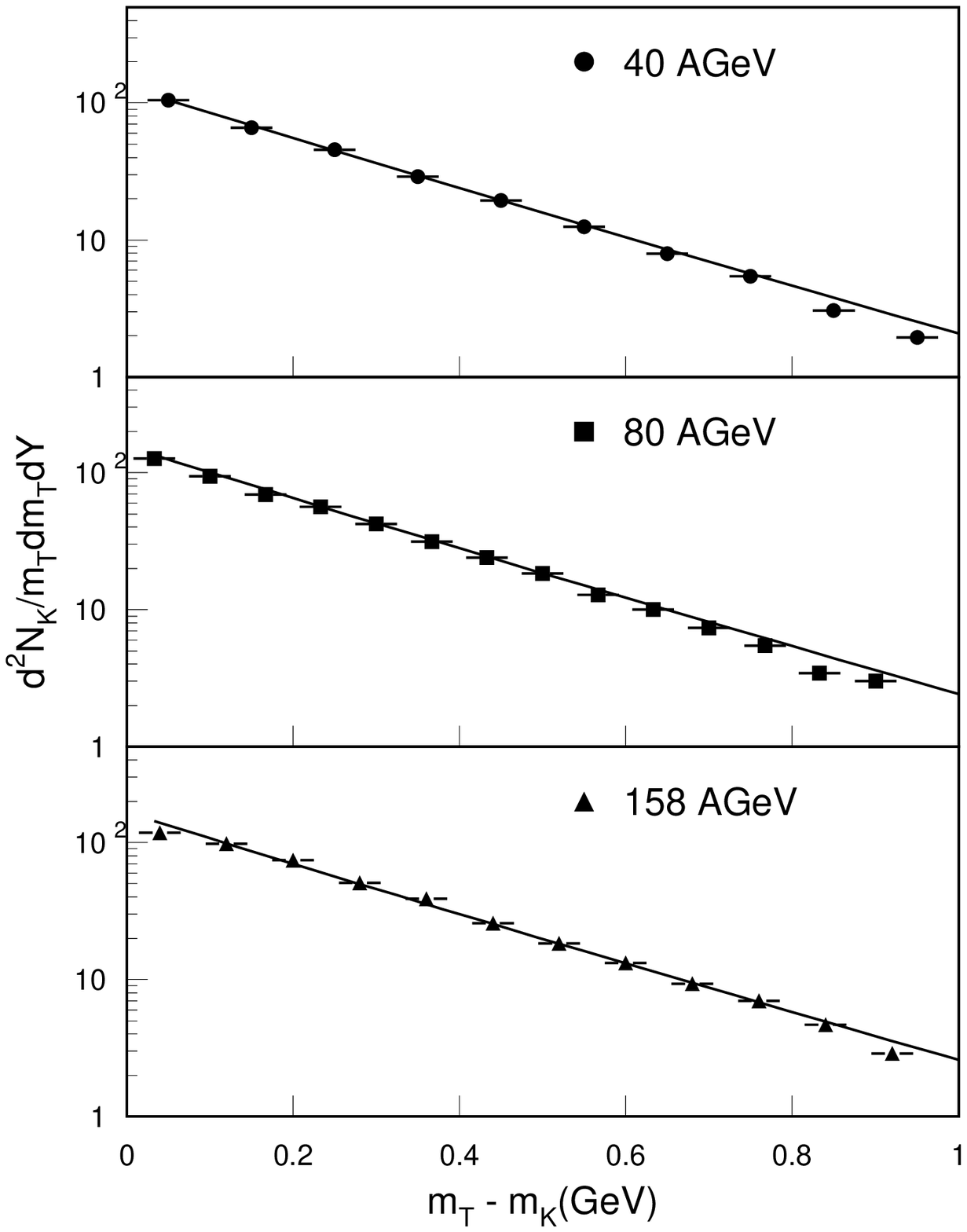}
\includegraphics[scale=0.4]{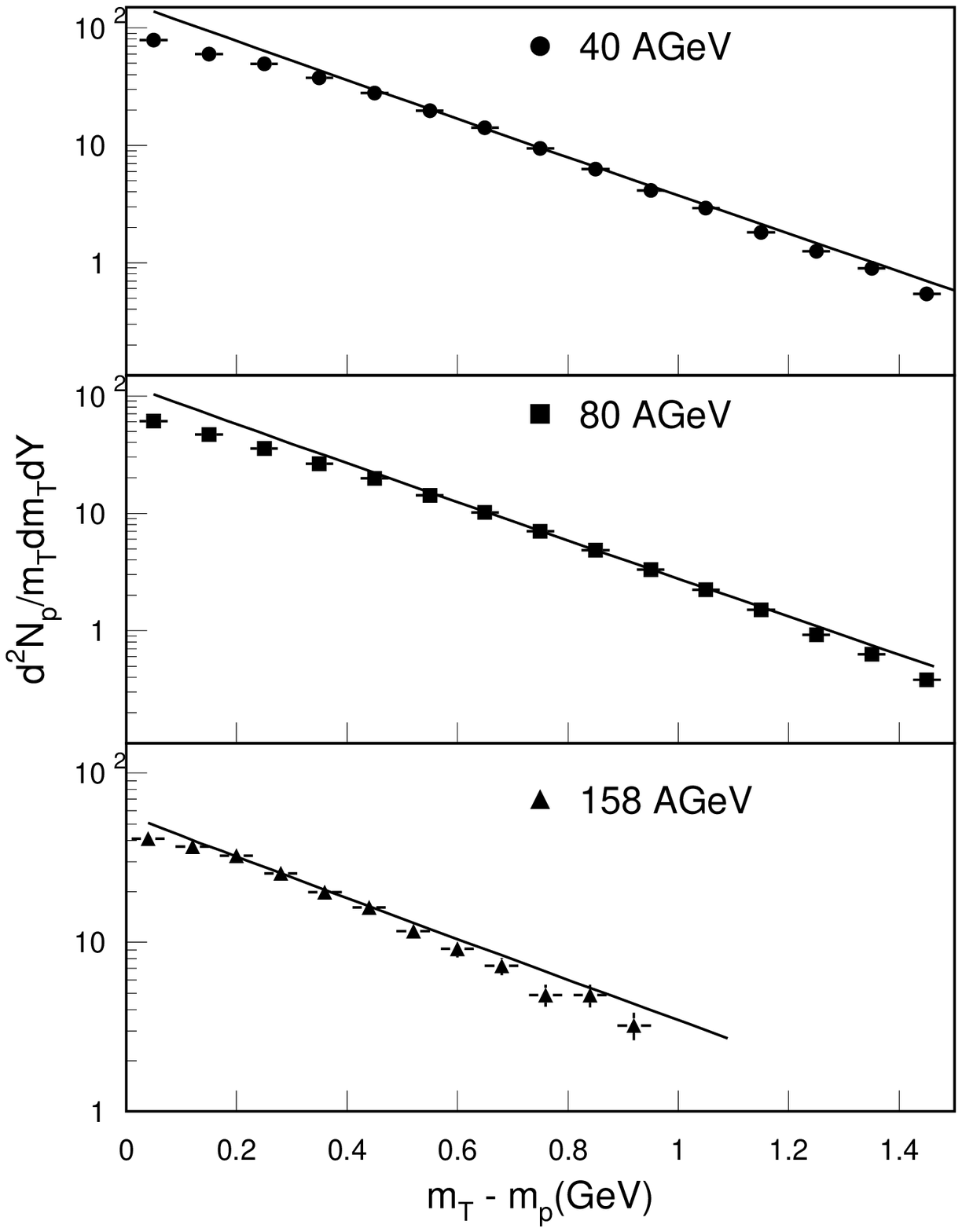}
\caption{
Transverse mass spectra for kaons and protons at SPS
energies. The solid line corresponds to results obtained from random
walk model. The model parameters T and $\delta$ are not fixed from pp 
and pA collisions respectively but are adjusted to give the best possible
agreement with the experimental data.
}
\label{arb_spectra}
\end{center}
\end{figure}
The results are shown in Fig.~\ref{arb_spectra}. 
One observes that the kaon spectra for the three different beam energies
are well explained by the model taking a common temperature of 235 MeV
and $\delta_{PbPb}$ of 0.25. The $\chi^{2}/ndf$ is less than 1.0 for 
all the three cases. It may be mentioned that this temperature although
very high compared to that obtained from pp (section III), is very close
to the temperature obtained taking radial flow into account~\cite{na49}. 
For the protons
the spectra is satisfactorily reproduced over a large range of transverse
momentum. For this we needed a common $\delta_{PbPb}$ values of 0.15 but
the temperature for 40 and 80 AGeV is 260 MeV, while that for 158 AGeV
energy it is 360 MeV.

\subsection{Transverse flow}
\begin{figure}
\begin{center}
\includegraphics[scale=0.4]{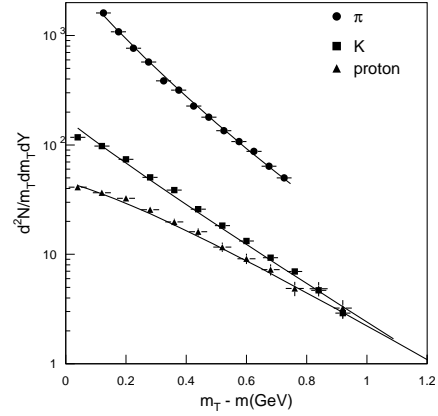}
\caption{
Transverse mass spectra for pions, kaons and protons at 158 AGeV Pb+Pb
Collisions. The solid line corresponds to results obtained from a 
combination random walk model and transverse flow.
The model parameters T and $\delta$ are fixed from pp 
p+Pb collisions respectively and the transverse flow velocity $\beta_T$ 
is taken to be $\sim$ 0.45. The normalization is adjusted to give the 
best possible agreement with the experimental data.
}
\label{aa_flow_spectra}
\end{center}
\end{figure}
The failure of random walk model to fully explain the transverse momentum
spectra of produced hadrons indicate that initial state $p_T$ broadening
alone cannot account for experimentally measured transverse mass spectra. One
observes that the model fails to explain the lower momentum part for kaons
and protons. This may be attributed to the lack of chemical equilibrium, which
can be accounted for by introducing a chemical potential for the hadrons
But the formulation of the model is such
that it affects only the normalization and not the slope. The other possibility
is that, there may be both initial state $p_T$ broadening and collective effect
like transverse flow. In other words the resultant spectra of produced hadrons
is the effect of both the above processes. In order to look at this 
possibility, we express the transverse mass spectra of produced hadrons as;
\begin{widetext}
\begin{equation}
\left( {dN\over dy m_Tdm_T}\right)_{total}
= C_1 \left( {dN\over dy m_Tdm_T}\right)_{random} + 
C_2 \left( {dN\over dy m_Tdm_T}\right)_{flow}
\label{eq8}
\end{equation}
\end{widetext}
where, the contribution from flow~\cite{flow,na49} is defined as,
$\left( {dN\over dy m_Tdm_T}\right)_{flow}
\sim m_TI_0 \left( {p_T\sinh\rho\over T} \right)
          K_1 \left( {m_T\cosh\rho\over T} \right)$, mean transverse velocity $\beta_T$ is defined through the relation $\rho~= atanh \beta_T$ and T is the 
freeze-out temperature here taken to be 150 MeV as seen from the slope of
pp spectra.

We tried to see if such a parametrization can explain the SPS data at 158 AGeV
beam energy and get the values of $C_1$ and $C_2$ which will tells 
the relative contribution of the two effects. 
The results are shown in the Fig.~\ref{aa_flow_spectra}. Since 
pion data is very well explained by the random walk model, we tried not to
fit it with the above parametrization. 
The aim here is to see how much contribution of transverse flow
in addition to initial state $p_T$ broadening through the random walk model 
can explain the data. 
From the figure we find that the 
results for the kaons and protons shows that such a parametrization works.
We have taken a value of $\beta_T$ of 0.45 for kaons and 0.48 for protons.
The values of $C_1$  and $C_2$ are 0.5 for kaons and protons, $C_1$ = 0.2
and $C_2$ = 0.8. The results indicate the flow effects dominates with 
increase in hadron mass.

\section{SUMMARY}
In summary, a systematic study has been carried out to understand the
transverse mass spectra of the produced hadrons in nucleus-nucleus
collisions within the framework of a simple random walk model.
The model is based on the assumption that the nucleus-nucleus collision 
is a superposition of nucleon-nucleon collision where  for
each successive interactions of a nuclear collision one creates a fire ball. 
The temperature of the fire ball is fixed from the available nucleon-nucleon
collision data. There is a  gain in transverse momentum through successive 
collisions, the propagation of which is assumed to follow a 
random walk pattern. The average transverse rapidity shift per collision or
the gain in transverse momentum per collision is obtained from the 
nucleon-nucleus collision data. Having fixed the two parameters of the
model, we then apply it to nucleus-nucleus collisions. It is observed
that, although the model fairly well explains the transverse mass spectra
of pions at SPS energies, it fails to do so for the higher mass hadrons like
kaons and protons. We find the choice of different distribution of random 
walk pattern yields the same result. 
This indicates the presence of true collective effects
in the nucleus-nucleus collisions. However it seems a considerable portion
of the gain in transverse momentum comes from initial state effects.
Considering both the initial state $p_T$ broadening and the transverse
flow effect, we are able to explain the SPS data.
We find that the contribution of transverse flow increases with increase
in hadron mass.
The avability of systematic data from pp to pA to AA collision at a 
fixed CMS energy would help in quantifying the effect of initial state
broadening of transverse momentum and hence the true amount of collective
effect.

\acknowledgments{I am grateful to the Board of Research
on Nuclear Science and Department of Atomic Energy, 
Government of India for financial support. I would like to 
thank Micheal Murray and Subrata Bhattacharyya 
for providing me the experimental data.
I would like to thank Jan-e Alam for critical reading of the
manuscript and for many helpful discussions.
}


\normalsize

\begin{thebibliography}{99}

\bibitem{tflow} I.G. Bearden et al. (NA44 collaboration), 
Phys. Rev. Letters {\bf 78}, 2080 (1997).

\bibitem{bdm} B. Mohanty, J. Alam, S. Sarkar, T.K. Nayak and 
B.K. Nandi, nucl-th/0304023. 

\bibitem{heinz}
U. Heinz, K.S. Lee and E. Schnedermann,
in ``Quark-Gluon  Plasma'', pp. 471,
Ed. R.C.  Hwa, World Scientific, Singapore (1992);
Jean-Paul Blaizot and Jean-Yves Ollitrault,
Adv. Ser. Direct. High Energy Phys. {\bf 6}, 393 (1990).



\bibitem{radial} C.M. Hung and Edward V. Shuryak,
Phys. Rev. C {\bf 57}, 1891 (1998).


\bibitem{cronin} J. W. Cronin et al., Phys. Rev. Lett. {\bf 31}, 1426 (1973);
                 J. W. Cronin et al., Phys. Rev. D {\bf 11}, 3105 (1975).


\bibitem{leonidov} A. Leonidov, M.Nardi and H. Satz, 
Nucl. Phys. A {\bf 610}, 124c (1996)
and  Z. Phys. C {\bf 74}, 535 (1997).

\bibitem{satz} H.~Satz, Proceedings of the International Conference on the
Physics and Astrophysics of the Quark-Gluon Plasma (ICPA-QGP'97), Jaipur,
India, March 15-21, 1997; hep-ph/9706342.

\bibitem{jane} Jan-e Alam, J. Cleymans, K. Redlich and H. Satz,
               nucl-th/9707042.


\bibitem{becattini}  F. Becattini,
Z. f. Physik C {\bf 69}, 485 (1996); 
F. Becattini and U.~Heinz,
Z. f. Physik C {\bf 76}, 269 (1997).



\bibitem{isr} B. Alper et al., Nucl. Phys. B {\bf 87}, 19 (1987)
and Nucl. Phys. B {\bf 100}, 237 (1975).

\bibitem{e802} E-802 Collaboration, T. Abbott et al., 
Phys. Rev. D {\bf 45}, 3906 (1992).

\bibitem{na44} NA44 Collaboration, I.G. Bearden et al.,
Phys. Rev. C {\bf 57}, 837 (1998); H. Boggild et al.,
Phys. Rev. C {\bf 59}, 328 (1999).

\bibitem{na49} NA49 Collaboration, S.V. Afanasiev, et al.
Phys. Rev. C {\bf 66}, 054902 (2002);
M. van Leeuwen et. al. (NA49 collaboration), nucl-ex/0208014.

\bibitem{flow} E. Schnedermann, J. Sollfrank and U. Heinz, 
Phys. Rev. C {\bf 48}, 2462 (1993).

\end{thebibliography}
\end{document}